# Concurrent codes: A holographic-type encoding robust against noise and loss.

David M. Benton

Aston Institute of Photonic Technologies, Aston University, Aston Triangle, Birmingham, UK. B4 7ET

**Abstract.**
Concurrent coding is an encoding scheme with 'holographic' type properties that are shown here to be robust against a significant amount of noise and signal loss. This single encoding scheme is able to correct for random errors and burst errors simultaneously, but does not rely on cyclic codes.  A simple and practical scheme has been tested that displays perfect decoding when the signal to noise ratio is of order -18dB. The same scheme also displays perfect reconstruction when a contiguous block of 40% of the transmission is missing. In addition this scheme is 50% more efficient in terms of transmitted power requirements than equivalent cyclic codes. A simple model is presented that describes the process of decoding and can determine the computational load that would be expected, as well as describing the critical levels of noise and missing data at which false messages begin to be generated.

## Introduction

Building robustness into data transmission is necessary and well established with methods of forward error correction using block codes, which are processed on a block-by-block basis, including (see for example [1][2]) cyclic codes, the Golay code, BCH codes, Reed Solomon codes and Hamming codes. Convolutional codes are processed on a bit-by-bit basis and include turbo coding[3] and Viterbi coding[4] which allows optimal decoding. These techniques generally operate on binary symmetric channels and parity information regarding neighbouring bits in the original data is encoded into a data stream, allowing errors in individual bits to be corrected. These codes are intended to efficiently correct random errors. Burst errors are non-random blocks of erroneous or missing bit values which are not efficiently dealt with using conventional random error correcting codes[1][5][6]. Cyclic codes such as Fire codes [7], and Reed Solomon codes[8][9] can recover corrupted symbols and provide burst error correction. The most common approach is to use interleaving. This distributes neighbouring bits into an array structure and deals with burst errors by redistributing the block of errors into individual isolated error events that can be dealt with by the random error correction code. Corruption also occurs due to interference from sources such as multiple user access, inter symbol interference, cross antenna interference and jamming. A variety of spread spectrum techniques[9][10] such as Code division multiple access (CDMA), Time division multiple access (TDMA), and Orthogonal Frequency Division Multiplexing (OFDM) are used to overcome interference and allow multiple user access.  Thus multiple layers of encoding are typically used to overcome different types of data corruption.

Concurrent coding is a little known technique developed by Baird, Bahn and Collins (BBC) [10]-[15] that was devised to offer a means of information encoding that could be resistant to the effects of jamming (intentional or accidental) without the need for the communicating parties to encrypt their communications with a shared key, such as using CDMA. The technique is unusual in using an asymmetric binary channel. This uses indelible marks representing 1's that are placed into a communication channel, with the absence of a mark representing a zero. The indelible mark is typically given by the presence of energy in a time or frequency channel. The indelibility provides the asymmetry as marks can be inserted ($0 \rightarrow 1$) but cannot be deleted ($1 \rightarrow 1$). Binary coded symbols, referred to herein as messages, are dissected, encoded and distributed throughout a much larger transmission. The name concurrent code arises from the ability to superimpose codes through a logical OR process. Many such concurrently coded messages can be overlaid into the transmission with the result being that each message is 'holographically' encoded into the transmitted codeword. This is to say that, broadly speaking, each part of the transmission codeword contains relationships with every message that is encoded within it. In the field of optical science it is appreciated that each fragment of a hologram can reproduce a version of the entire hologram from which it came (albeit at reduced quality). It is therefore reasonable to speculate that a holographic encoding technique might exhibit similar qualities, such that missing parts of a transmission could be reconstructed.

Resistance to jamming arises by requiring a jammer to expend large amounts of energy across frequency and/ or time bins in an effort to hinder communications. Whilst jamming is an important security concern [16], the emphasis in this paper is more general. The original concurrent coding investigated the effects of the addition of false marks through jamming or noise. The concept of indelibility refers to the inability of an attacker to deceive the receiver into interpreting a 1 as a 0 through the use of tailored emissions. However energy can be physically blocked thus effectively changing 1's to 0's. As this is more difficult to control and would be slow to implement in a deliberate fashion, it was not a plausible jamming technique and hence was not considered. Indeed it was stated that the probability of a $1 \rightarrow 0$ occurrence must be driven to zero[17]. Slow physical signal blockage will result in bursts of 0's and not in individual bit reversals. The distribution of individual message bits around the codeword in a manner similar to random interleaving, allows the concept of mark indelibility to be maintained and burst errors can be treated independently.

In this paper the ability of concurrent coding to correct for burst errors as well as random errors has been demonstrated for the first time. Concurrent coding is a different approach to most conventional ways of encoding and can provide a robustness against loss of information in a number of ways. It is helpful to provide a list of characteristics of concurrent coding that can be held against conventional symmetric approaches.

- An asymmetric encoding method that utilises indelible marks means that once data is encoded into a transmission it cannot be removed other than by physical blockage of a large majority of the transmitted signal. Thus once a 'message' is sent it can always be received.
- Messages cannot be corrupted, only obscured by the decoding of additional false messages called hallucinations.
- Concurrent coding corrects for both random and burst errors simultaneously with a single stage of encoding. A conventional approach to overcoming burst errors, random errors and interference might having the following possible flow:

    Data → Interleaving → Parity Encoding → Spread Spectrum → Transmission.

In contrast, as we shall see, the equivalent process for concurrent coding is:

Data → Concurrent Coding →Transmission.

- Concurrent coding does not use cyclic coding and does not generate a syndrome.
- Marks can be shared between encoded messages leading to more efficient use of transmitted energy
- Concurrent coding is significantly simpler to understand than cyclic coding schemes.

The requirement for indelible marks limits the modulation schemes able to carry concurrent coding to binary formats therefore symbolic formats cannot be used as one symbol can be transmuted into a different one. Schemes such as on-off keying are particularly well suited but others such as amplitude shift keying (ASK) and frequency shift keying (FSK) are also viable.

In this paper the principles of concurrent codes are described followed by an implementation that reveals the characteristics of this encoding scheme with relation to noise and missing data. Comparisons are made with an interleaved Hamming code. Following this a model is developed and its expectations are compared with the outcomes from the implementation. Finally a description of applications and extensions is presented. The motivation behind this paper is to highlight some of the unique characteristics that concurrent codes can offer into a vast field of encoding techniques that are dominated by symmetric formats, with the view that some less conventional usages could benefit from this approach.

## Method

### The Principle of Concurrent Coding

The principle behind concurrent coding is well described in [10][12][13] but briefly described here. Concurrent coding uses the unique linear sequence of 0's and 1's in a message word to generate a pattern of 0's and 1's uniquely distributed across a larger codeword space. A message is broken down into linearly expanding sub-sequences of bits – pre-fixes starting from the least significant bit and incrementally increasing in length. Each pre-fix is then passed through a hashing function. The output of the hashing function is used as the address of a mark to be placed in the codeword space. As a simplistic example the message 1101 will produce addresses from the arbitrary hash sequences H(1), H(01), H(101), and H(1101). Multiple messages can be combined via an OR process into a single codeword before transmission, as shown schematically in Figure 1.

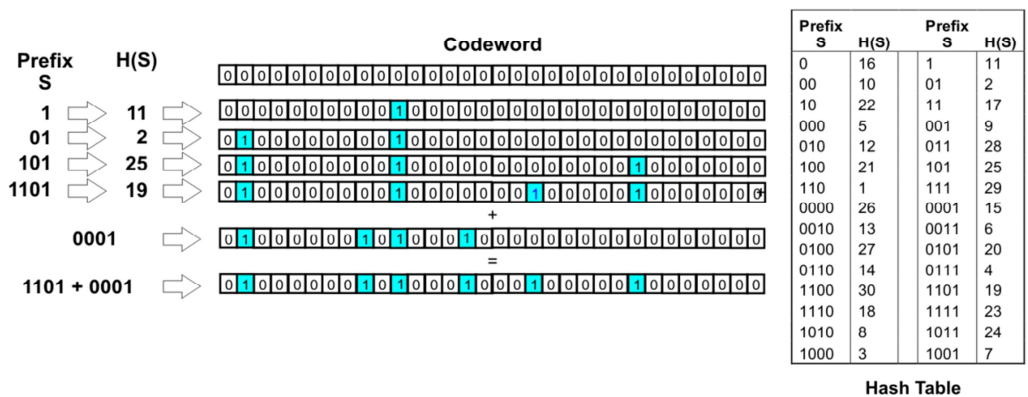

Figure 1. A schematic diagram of producing codewords from messages. The 4 digit message 1101 is sequentially hashed using the hash table and marks placed in the codeword space. The process is repeated for a second message 0001. The two results can be ORed together to produce a final packet of messages. Notice that the final packet contains six 1's representing the eight bits of the two input messages. This arises from the two prefixes '1' and '01' , common to both messages sharing the same marks at positions 11 and 2.

The Hash function distributes the message bits around the transmission codeword. This performs a similar function to that of random interleaving which provides resistance to burst errors and can be used for multi user access [18][19][20]. However there is a fundamental difference here. Interleaving rearranges a fixed number of bits into a different order, whereas concurrent coding can map multiple prefixes from different messages onto the same final mark in the codeword.

The decoding of the message proceeds by trying the values 0 and 1, (the first potential message bits) and passing them through the hashing function, then examining the received codeword. If a mark is found at the position indicated by the output of the hash functions then the message value is retained for further analysis. If no mark is found all possibilities with the input sequence cannot be found and analysis not pursued. For the retained values the next step in the sequence is examined with both 0 and 1 appended i.e. if H(1) found, next step is H(01) and H(11), again retaining those attempts that result in an associated mark present in the codeword. The process is repeated for the number of bits in each message. The process forms a decoding tree as represented in Figure 2.

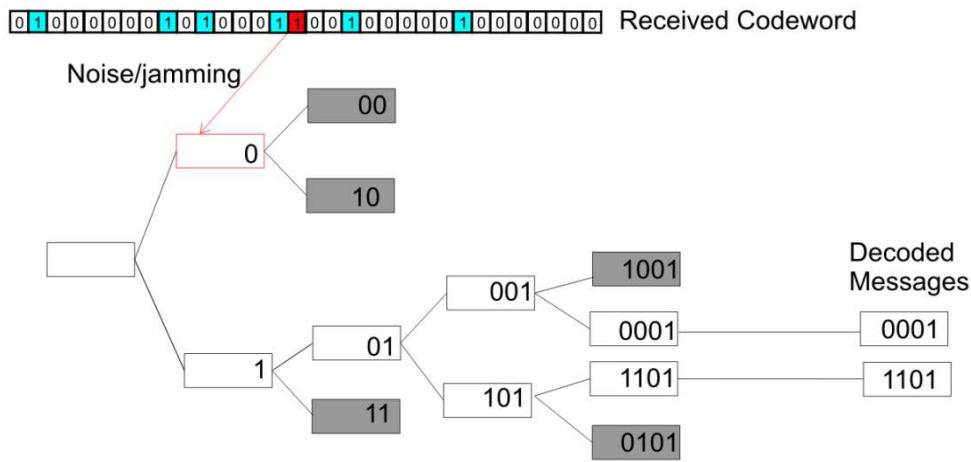

Figure 2. A representation of the decoding tree. Grey boxes represent dead branches where no corresponding mark was found in the codeword

The effect of noise or jamming that adds marks into the codeword is also shown by the highlighted bit in the codeword. It can be seen that whilst initially extra branches are retained they are quickly lost as no marks corresponding to genuine messages are found.

With a significant level of noise there will be routes through the decoding tree that result in messages not actually present in the original codeword. These false messages are referred to as hallucinations. To help reduce hallucinations a number of constant value, (0) checksum bits are appended to each message (e.g 1101 becomes 001101).

Conventional error correction schemes use separate encoding processes to embed data relationships between bits in the form of parity information or cyclic codes (to correct for bit errors) followed by a protection against burst errors with interleaving, the classic example being the encoding used in reading CDs[8]. The decoding stage reverses this by de-interleaving followed by a check and correction for errors. In many cases resilience to interference is ensured by adding a spread spectrum stage to the interleaved and encoded data. Concurrent coding does not separate these stages as they are all inherent in the process. Decoding does not have any check for errors, as genuine messages cannot be removed or corrupted, thus genuine messages are always decoded correctly, but any false messages cannot be distinguished. However the receiver knows that the decoded message set always contains the original message set as a subset. It is notable that any codeword can only uniquely decode 1 version of a message and the order in which messages are decoded need not have any relation to the intended order of the message set. Hence the messages themselves must have some mutually understood independent meaning, such as being indices within a larger codebook. Concurrent coding is then quite different from conventional methods used to incorporate robustness into a communication channel.

## The Hash Function

The hash function used in this technique to determines the addresses at which marks are placed into the codeword. This function can be called many times, particularly in the decoding tree, for which the processing requirements can be a limiting factor in the performance and could therefore limit bandwidth and usability. Indeed BBC recognised the impact of the hash function and developed the Inchworm Hash to speed up the process [14] and subsequently the Glowworm Hash [21]. Simplicity and speed of implementation are important factors in translating concepts into viable technologies. The driving principle of the hash function is redistribution around the codeword, not in this first instance of providing security in the encoding. Therefore the most important consideration in the present context is that the hash functions distributes marks uniformly throughout the codeword and does not produce significant addressing clashes leading to ambiguous decoding. For this purpose a pseudo random bit sequence (PRBS) generator was found to be a suitable and simple variant of the hashing function with the added attraction of simplicity of implementation.

An 11 bit PRBS addresses a codeword space of 2048 bits and enables 8 bit messages with 2 check sum bits to be used. A PRBS is attractive because it can be implemented in either hardware or software without requiring a large number of operations. The PRBS used in this work is represented in Figure 3. Whilst this approach does not hash pre-fixes as presented earlier, the progression through each bit of the message produces an output state dependent upon the current bit and all the previous bits. The output of each cycle is used as the address for placing a mark in the codeword.

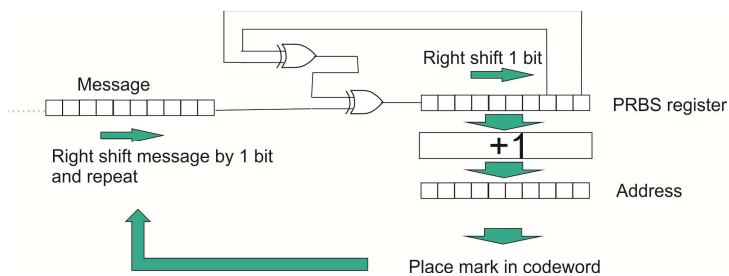

**Figure 3. A representation of the 11 bit PRBS used for hashing.**

## Implementation

A concurrent code model was coded using LabView software with a view to exploring the performance of this technique with a variety of data conditions and identifying strengths and weaknesses. A word about the scale. In this work simplicity and practicality are driving principles behind the design. For this reason the codeword length was kept relatively short at $2^{11}$ bits, allowing 8 bit messages to be encoded (convenient for transmitting ASCII codes for example) and equally allowing a simple PRBS to be used. In addition 2 check sum bits were appended resulting in 10 marks being required for each encoded message. The BBC implementations often focussed on much longer codewords with individual messages as long as 200 bits as well as more involved hashing algorithms. One downside of this is significant latency in the final decoding of the transmission; an issue that is sometimes encountered with interleaving where the effect is much reduced in comparison to concurrent coding. For concurrent coding the entire codeword must be received before the message decoding process can begin, followed by what could be a lengthy computationally involved decoding process. The opposite approach is chosen here, to keep the latency down with short codewords and to reduce computational load with simple hashes.

# Results

## The Effect of Noise

A chosen number of randomly generated 8 bit messages were passed through a concurrent coding algorithm and a codeword prepared from the overlaying of all the encoded messages. This prepared codeword was then degraded through the addition of a controllable level of marks randomly placed into the codeword. An important characteristic of concurrent coding is that genuine messages placed into the codeword cannot be deleted, they can only be obscured by the presence of hallucinations, a feature arising from the indelible marks used and the binary asymmetry. Thus the level of hallucinations generated is a critical parameter in assessing the effective 'information to noise' level. At this stage attention is drawn to the obvious situation where determining if a mark is present requires assessing if the signal exceeds a threshold level. Once the signal exceeds the threshold its absolute amplitude is irrelevant. Hence there is equivalence between random marks generated by a noisy signal (low threshold) and a strong signal with random marks from, say, interference or jamming. Consequently we shall deal with the binary signal composed of genuine and random marks and not be concerned with how that situation arose. From this point the term 'noise' refers to randomly placed marks in the codeword which were not originally encoded.

To provide a comparison with conventional techniques an 'equivalent' encoding system was also implemented. Direct comparison with conventional techniques is not possible because of the asymmetric nature of concurrent codes. Errors in decoding arise as false decodings but do not corrupt genuine messages, whereas in conventional approaches genuine messages are lost if the noise/error level exceeds the capability of the encoding to correct for them. Providing 'equivalence' requires taking the same size of messages and encoding to correct for transmission errors with a codeword of the same length. In this case 8 bit messages were halved and encoded with an H(8,4) Hamming code. The results were then interleaved into a 2048 bit codeword space. The prepared codeword was then subjected to noise and degradation before being decoded.

When adding noise to into the codeword a standardisation was used where the number of marks present for a single message – in this case 10 marks - was defined as 0dB. A fixed number of messages, whose contents were generated randomly, were encoded and then random marks of a chosen scale (in dB) were added into the codeword. This was then decoded and the level of hallucinations(errors) recorded. This process was repeated to produce an average level of hallucinations(errors). Figure 4 plots the number of hallucinations produced as the noise level is increased, with a fixed number of 10 messages encoded. Hallucinations start to appear at a noise level around 18dB. The generation of hallucinations was found to be dependent upon the initial state of the PRBS register and hence values that generated no additional hallucinations at lower noise levels were used.

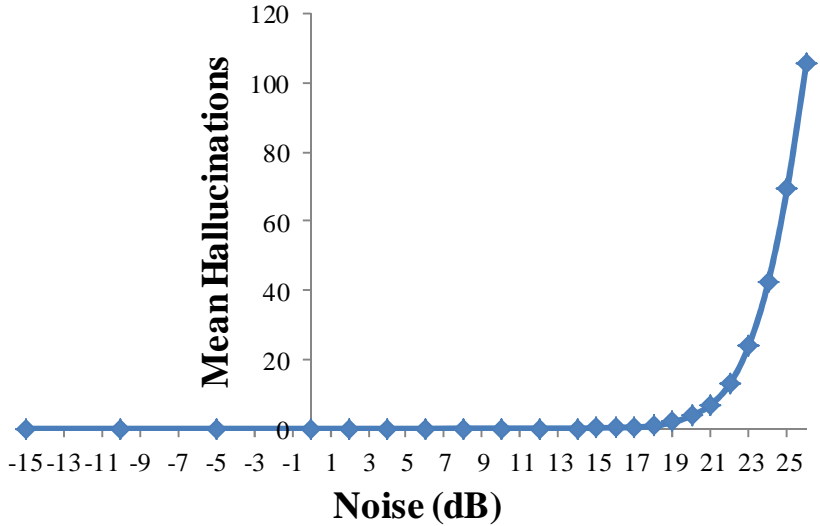

**Figure 4.** The number of hallucinations vs the level of random noise marks introduced into the codeword. Results are from an encoding simulation.

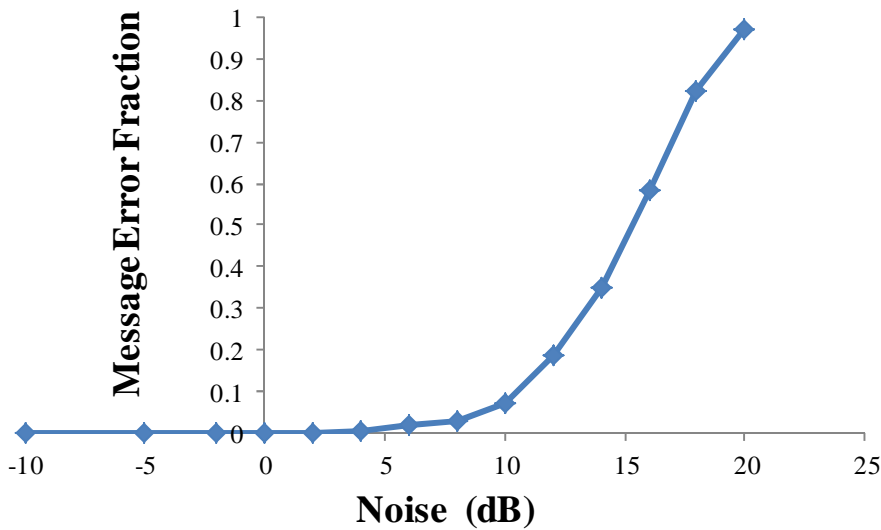

**Figure 5.** The effect of random noise upon the Interleaved Hamming encoding. The message error fraction is the number of genuine messages that are incorrectly decoded relative to the original messages.

The effect of the number of messages encoded was investigated and found that, even with some noise introduced, no hallucinations were created with 80 messages encoded, although the decoding time increased significantly. In comparison the Hamming interleaved code, with the addition of (binary symmetric) noise, behaved as shown in Figure 5. The error fraction, averaged over multiple repetitions, is the ratio of errors that occur in the original 10 messages (not in the empty message slots). It can be seen that errors start to appear at noise levels above 4dB.

**Calculating the Number of Marks**

For an N-bit codeword there are $2^N$ possible unique combinations. As unique messages are added to the codeword the number of combinations available is reduced as $2^N-m$, where $m$ is the number of messages. The number of marks common between messages is dependent upon the matching pre-fixes (starting from bit index zero) between different messages. For a population of *m* messages there will be a minimum number of pre-fix matches within the population, given by *floor(log$_2$(m))*, where floor represents the largest integer value less than the argument. Writing the number of messages as:

$m = 2^a$, *a* is the effective number of bits in common between messages and therefore represents the number of post-hash marks shared by those messages

$$a = \log_2 m \tag{1}$$

The number of marks produced for *m* messages is then

$$Z(m) = Nm - m \log_2 m \tag{2}$$

The average number of marks produced for a given number of randomly generated messages was recorded. This data is shown in Figure 6 with up to 100 included messages. The estimated number of marks is also shown. This estimate displays the trend of the number of marks but consistently underestimates the number of marks produced by more than the standard deviation of the measurements –typically 2 standard deviations for *m*>10. This discrepancy is most likely due to properties of the PRBS hash function and requires further investigation but does adequately describe the trend of mark generation.

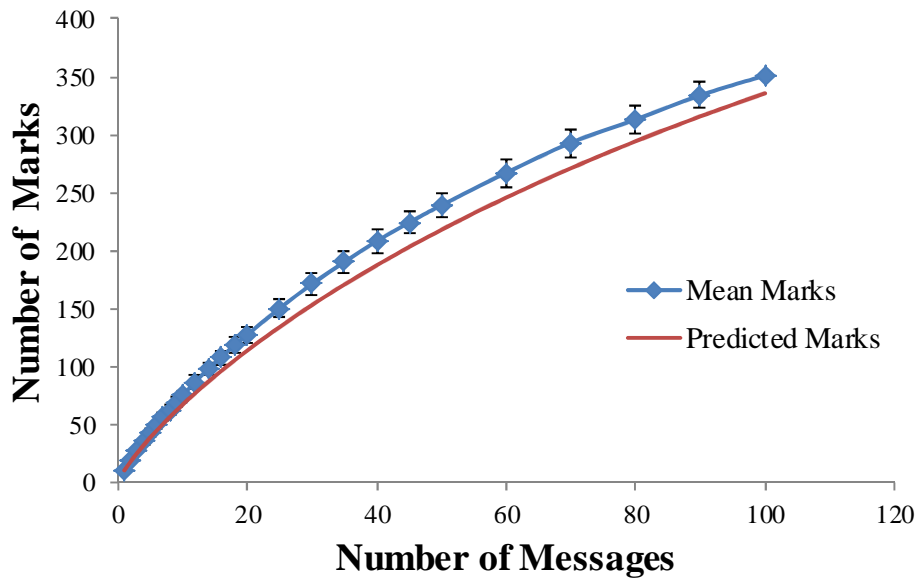

Figure 6. The number of marks in the codeword determined from an encoded model and predicted (using equation 2) for a given number of included messages.

Given the number of messages, the number of marks can be determined and hence the 'message signal' level can be determined. Clearly introducing more messages improves the 'message signal' to noise ratio, but does so at the expense of speed of decoding. It is also worth indicating that when the codeword contains a large fraction of message marks, the relative effect of any noise is enhanced because a single mark can be associated to many potential message marks. This will add significantly to the processing overhead of the decoding and increase the number of hallucinations. Maximising the number of messages present in the codeword may not be the optimum strategy for guarding against the effect of noise or interference.

Knowing how many marks are produced from a given number of messages it is natural to consider the situation where a codeword is received without knowledge of how many messages it contains. The receiver would then wish to determine, given a number of marks received, how many messages should be expected. Solving equation (2) for the number of messages $m$ results in:

$$m = \frac{-Z \log 2}{W(-2^{-N} Z \log 2)} \qquad (4$$

Where $W$ is the Lambert W function which has to be evaluated explicitly through a series expansion of the form

$$W(x) = \sum_{n=1}^{\infty} \frac{(-1)^{n-1} n^{n-2}}{(n-1)!} x^n \qquad (5$$

which is only reliable for real values of x<0.4. In practice this estimation is simpler when a cubic polynomial is fitted to the data of Figure 6. Calculating the number of expected messages is really only effective as a check that the decoding process has produced sufficient messages or indeed too many messages.

## Effects of Burst Error Intermittency

Intermittency in a signal can arise due to misalignment of directed beams, environmental effects such as scintillation or fading. Removal of a large contiguous chunk of data can be catastrophic for most encoding methods, hence the addition of interleaving. To test the earlier assertion that missing information can be reconstructed it needs to be first established that data can be identified as missing. Where few messages are present in a codeword there will naturally be large gaps between marks. For a given number of messages the probability of a gap is the probability of a series of consecutive zeros and can be determined from the Poisson distribution on the assumption that the hash function distributes evenly throughout the codeword. The probability of finding a block of empty marks of length $E$ in a codeword of length $C$ is:

$$P_B = \frac{C-E}{C} P(0)^E \qquad 6)$$

Where $P(0)$ is the Poisson probability of an empty mark obtained from the general Poisson formula:

$$P(x) = \mu^x \frac{e^{-\mu}}{x!} \qquad 7)$$

The parameter $\mu$ is the mean value which is given by the mark density of the codeword. For a given number of messages, $m$, the mean value can be obtained from equation (4) as

$$\mu = Z(m)/C \qquad 8)$$

Figure 7. shows a histogram of the probability of a block of empty marks equivalent to 10% of the codeword size which varies as a function of the number of encoded messages. For only 5 messages this probability is 2% suggesting that missing data chunks can be reliably detected with a minimum of 5 messages. More messages makes smaller data gaps detectable, with a 5% gap having a probability of 0.3% with 20 messages encoded.

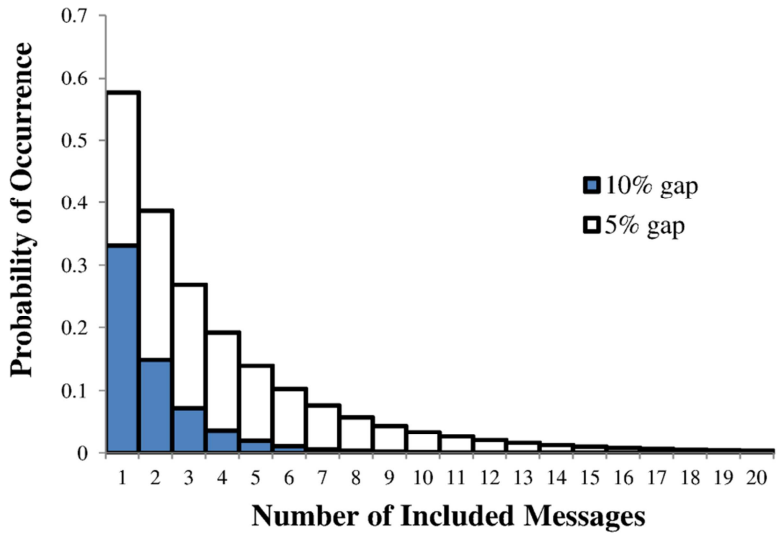

**Figure 7.** The probability of a block of empty marks equating to both 5% and 10% of the codeword size as a function of the number of encoded messages.

Data gaps were introduced into prepared codewords by zeroing contiguous chunks. In the decoding process received codewords were first examined to identify the presence of data gaps above a size threshold. Decoding branches whose marks would have appeared in the gaps were not discarded but retained and examined in the next decoding round. With a list of 30 random messages encoded, the decoding performance for different sizes of missing data chunks is shown in Figure 8. Perfect decoding with no hallucinations can be achieved with up to 40 % of the codeword missing. Hallucinations begin to appear at 40% missing and increase rapidly as more codeword is removed. However even with significantly more than 40% missing, the original messages can still be successfully received even though they are obscured by a large number of hallucinations. This is remarkable robustness.

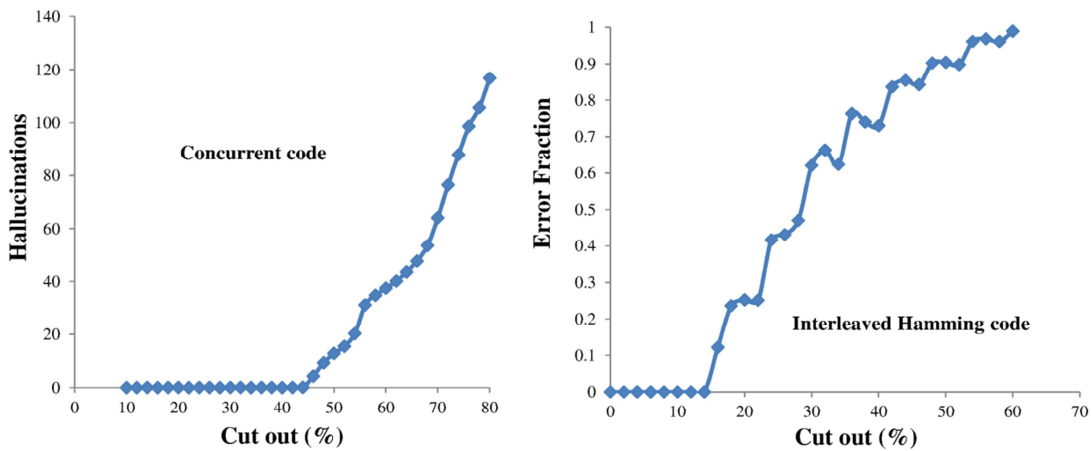

**Figure 8.** Decoding performance of a codeword containing 30 messages with various sizes of missing data cut out, shown as a percentage of the whole codeword. The left hand plot shows the hallucinations generated in the concurrent code scheme, the right hand plot shows the fraction of errors introduced into the original messages using the interleaved Hamming code.

**Comparison with Interleaving**

It has been shown that concurrent coding is able to correctly reconstruct data with 40% of the transmitted codeword missing and thus any equivalent conventional encoding scheme, involving additional interleaving, must have the capability to identify and correct a 40% BER. The depth and nature of the interleaving must be chosen to adequately compensate for the expected scale of burst errors, and therefore knowledge of the behaviour of the communication channel is required in advance. Details of the interleaving scheme for the HIC are given in Appendix 1.

The result of decoding performance with missing codeword chunks for the Hamming interleaver is shown in the right hand plot of Figure 8 where the fraction of errors in the decoded messages is given. It can clearly be seen that when the missing data cut out exceeds 2 interleaving spacings (12.5% of the codeword, see Appendix) errors start to appear, corresponding to exceeding the 2 bit correction capability of the Hamming code.

To summarise the comparison between concurrent codes and interleaved encoding, an interleaver has a maximum size of burst error that can be corrected, in this case 12.5% of the codeword after which messages become corrupted. In contrast concurrent coding has a minimum detectable burst error and knowledge of the number of messages being encoded is not needed, all messages are successfully decoded but can be obscured by hallucinations.

**Comments on Comparison with Reed Solomon encoding**

Reed Solomon (RS) encoding is an effective method for correcting symbol errors that can occur as a result of signal dropout [9], and hence some comparison with RS encoding is appropriate. The first comparison to draw is that concurrent codes only work with binary signals whereas RS codes operate with symbols and are therefore perhaps complimentary. RS encoding corrects for symbol errors irrespective of the number of bit errors in the symbol and it is the symbol error rate that determines the decoding fidelity,.

However whilst a 40% dropout can be successfully reconstructed using an RS code, any additional noise would lead to symbol errors, whereas with concurrent code this will lead to hallucinations in addition to the genuine messages.

The whole codeword space for RS encoding must be filled which is clearly not the case for concurrent coding as marks are only included representing messages. A concurrent codeword could therefore contain just a single message and contain only 10 marks. Concurrent codes are therefore more efficient in terms of transmitted energy content. Indeed comparing with the RS code, the equivalent concurrent code containing 100 messages would generate around 340 marks whereas the RS codeword would contain an equal number of 0's and 1's and therefore 512 marks. The RS code is therefore 50% less efficient than the equivalent concurrent code, which could be a significant benefit where low power usage is required.

## Modelling for Computational load

Choosing the number of messages to include in a concurrent coding will depend upon factors such as the need for overcoming intermittency (a minimum number), the time taken to decode the codeword which increases as the number of messages increases, and the effects of noise. A simple model for understanding the computational load can be understood as follows. Each round of decoding involves 2 calls to the hash function for every branch in the decoding tree that survives. The number of possible branches at each decoding round is $2^i$ where $i$ is the integer index of the decoding round and $1 \leq i \leq N$. Assuming for simplicity that whilst $2^i < m$ (the number of messages) all available branches are live, then when $2^i > m$ hallucinations are created through the presence of noise. Each live branch can spawn 2 branches in the next round until the check sum bits are reached in which case branches can only be killed and not created. The number of live branches $B$ at decoding round index $i$ is given by:

$$B_i = 2^i \text{ if } 2^i < m \qquad 9)$$

When the number of possible branches exceeds the number of messages present:

$$B_i = m + (2^i - m)P_n^{i-a+1} \text{ for } i>a \qquad 10)$$

Where $a=floor(log_2(m))$ and $P_n$ is the probability of each branch finding a mark arising from noise. This is given simply as the noise fraction, $n$, the ratio of the number of noise marks to the codeword length. $P_n = n = Noise/2^N$. When the decoding process reaches the checksum bits the hallucinations are killed off and the number of live branches is:

$$B_i = m + (2^b - m)P_n^{i-a+1} \text{ for } i>b \qquad 11)$$

Where $b$ is the index at which $k$ checksum bits are used, $b=N-k$.

In Figure 9 the left hand plot shows the number of live branches at each round of the decoding process for various numbers of encoded messages and a noise fraction of n= 0.45. The right hand plot in Figure 9 shows the number of branches at each decoding round with 32 encoded messages and various noise levels. It can be seen that as $2^i > m$ the number of branches decreases as long as $n<0.5$. After round 8 the checksum bits begin to kill branches. Note that this assumes a perfect hashing function with no clashes or interactions between messages. The computational load is simply the sum of the number of branches at each stage and is shown in
Figure 10 for two noise levels.

The number of expected hallucinations is then calculated as:

$$H = B_N - m \qquad 12)$$

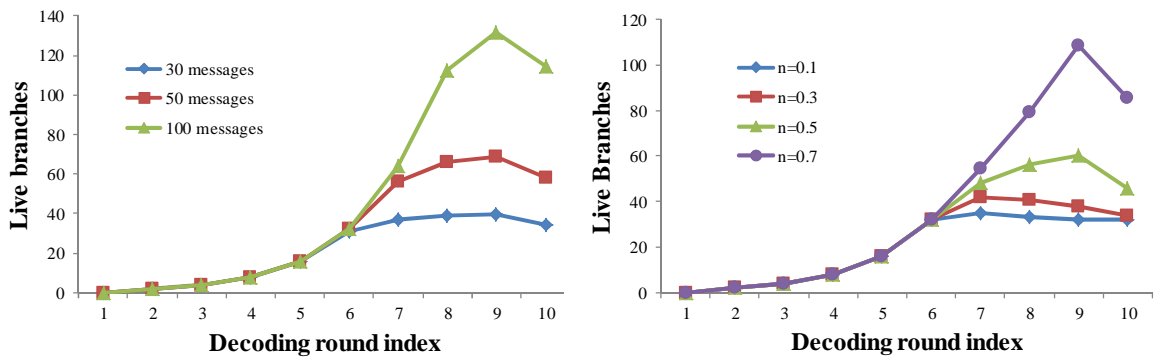

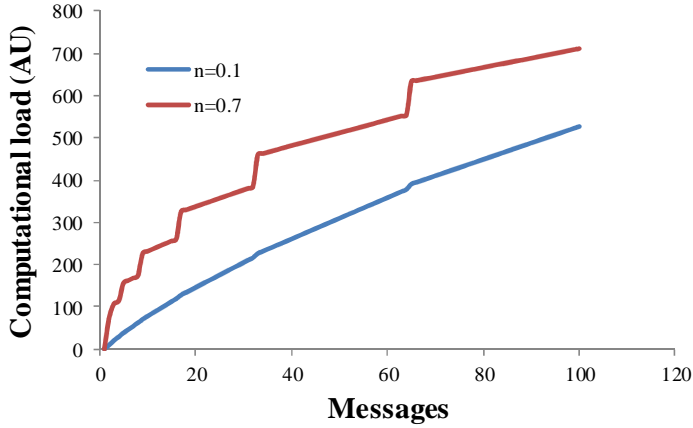

Figure 9. The number of live branches at each round of the decoding process. The left hand plot shows the numbers of live branches for various numbers of messages encoded and a noise fraction of 0.45. The right hand plot shows the effects of noise upon the number of branches with each having 32 messages present.

Figure 10. The computational load for modest and high noise levels

## Modelling for Noise and Missing Data

Accounting for the effect of missing data in the form of contiguous gaps can be done as follows. It is assumed that the perfect hash function distributes marks randomly and evenly throughout the codeword, with no order relating to the bit position. Any branch leading to a mark that falls into an identified gap will be retained in the decoding process. The probability of this, given the assumptions of a perfect hash, is the ratio of the gap extent to the codeword size. This probability can be included into equations (9) and (10) as

$$B_i = m + (2^i - m)(P_n + P_g)^{i-a+1} \text{ for i>a} \qquad 13)$$

$$B_i = m + (2^b - m)(P_n + P_g)^{i-a+1} \text{ for i>b} \qquad 14)$$

Where $P_g = g = gap/2^N$. Figure 11 shows a plot of the number of live branches for each decoding round for various sizes of gap present. The number of messages was kept constant at m=32 and the noise fraction was zero.

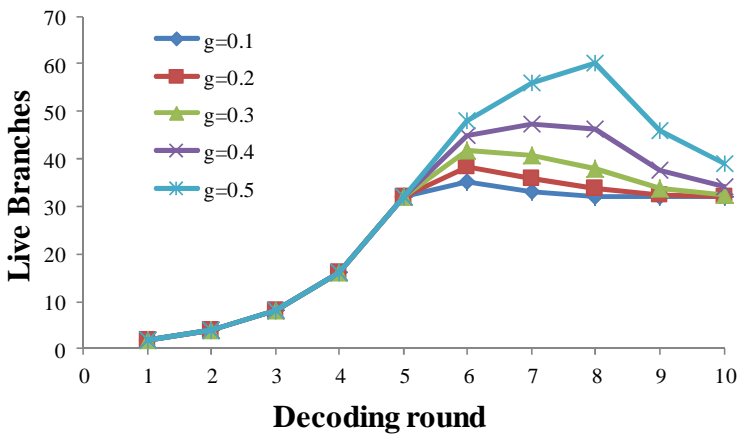

Figure 11. The number of live branches at each decoding round for differing gap size fractions. A constant number of message m=32 was used.

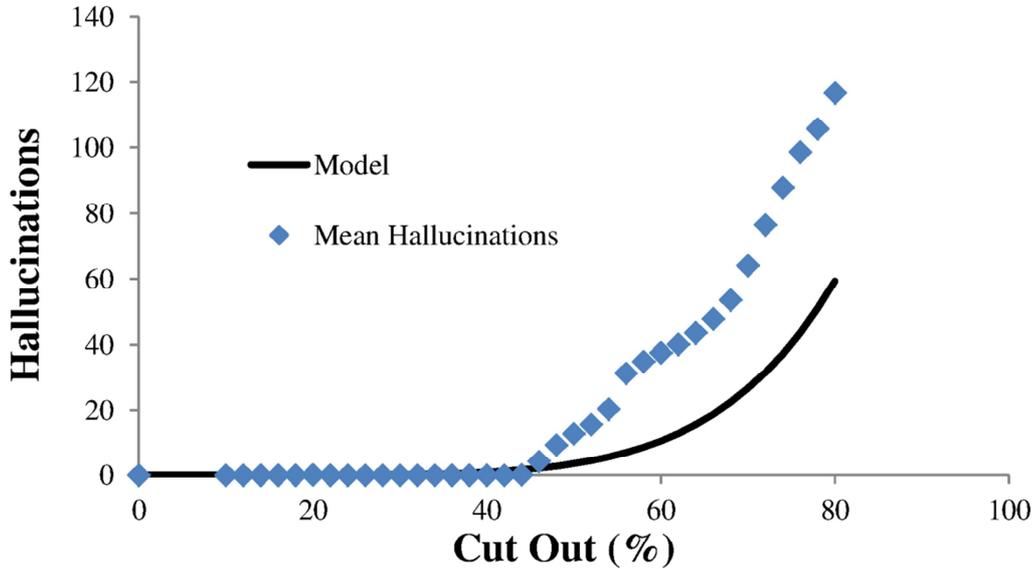

Figure 12. A comparison of the measured level of hallucinations with cut out size, in comparison to predicted hallucination level.

The predicted level of hallucinations is shown along with measured hallucination levels in Figure 12. At high levels of missing data this underestimates the number of hallucinations. This is most likely due to the PRBS algorithm being an imperfect hash function. Both distributions show good agreement in where hallucinations start to be produced. The threshold for hallucination production can be determined using equations (12) and (14) by setting $H=1$;

$$(n+g)_t = \left(\frac{1}{2^b - m}\right)^{\frac{1}{N-a+1}} \quad \quad 15)$$

Setting the noise $n=0$, the gap fraction threshold for hallucination production can be determined using values from Figure 12 with $m=32, N=10, a=5, b=8,$ this gives a threshold value of $g_t=0.4$, which is in good agreement with the data plotted in Figure 12.

Setting the gap fraction to zero the same calculation gives the threshold at which noise starts to generate hallucinations. This value of $n_t=0.4$ corresponds to 16dB, in reasonable agreement with the measured data plotted in Figure 4. It seems then that a concurrent code has an inherent tolerance for a combination of gaps and noise as both contributions are additive in their effect upon the probability of generating hallucinations.

The way that the total number of marks (messages plus noise) affects the production of hallucinations is seen from:

$$marks = 2^N (n+g)_t + mN - m\log_2 m \quad \quad 16)$$

$$marks = 2^N \left(\frac{1}{2^b - m}\right)^{\frac{1}{N-a+1}} + mN - m\log_2 m \qquad 17)$$

Plotting this relationship shows a curve representing the number of marks in the codeword at which hallucinations start to appear and this is shown in Figure 13. The *floor* function has been removed in the calculation of parameter *a* as this lead to large discontinuities as the number of messages crossed a power of 2.

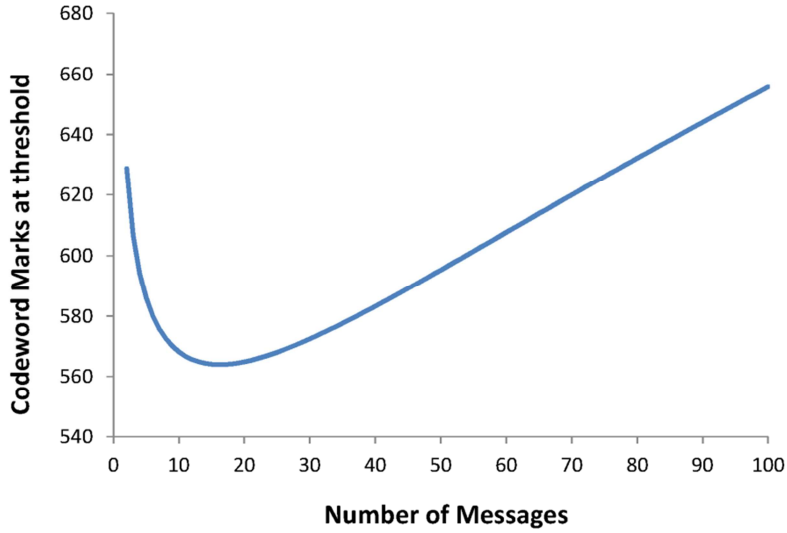

**Figure 13. The variation of the number of marks present in the codeword representing the threshold at which hallucinations are produced. This is for a codeword of 1024 bits with 8 bit messages.**

## Determining the Signal threshold

Hallucinations appearing at the end of the decoding tree are the equivalent of noise produced in the decoding. Having stated that the actual signal amplitude is irrelevant provided the mark amplitude is greater than a certain threshold level, we can now examine what that threshold level should be to avoid hallucinations.

Assuming that the noise is normally distributed, the probability of a noise value being below a value *x* is determined by the cumulative probability function.

$$F(x) = \frac{1}{2}\left[1 + erf\left(\frac{x - \mu}{\sqrt{2}\sigma}\right)\right]$$

16)

Where $\mu$ is the mean noise value, $\sigma$ is the distribution width and *erf* is the error function. The probability of the noise value exceeding the threshold $E_t$ is then

$$P(x > E_t) = 1 - F(E_t) = \tfrac{1}{2} - \tfrac{1}{2}erf\left(\frac{E_t - \mu}{\sqrt{2}\sigma}\right) \qquad 17)$$

The level of random marks that gives rise to hallucinations was determined previously in terms of a noise fraction, which is the probability of a random mark, we therefore equate 17) to $n_t$. The threshold level is then given by

$$E_t = \sqrt{2}\sigma erf^{-1}(1-2n_t) + \mu \qquad 18)$$

where $erf^{-1}$ is the inverse error function. The signal to noise level required is

$$\frac{E_t}{\mu} = \sqrt{2}\frac{\sigma}{\mu} erf^{-1}(1-2n_t) + 1 \qquad 20)$$

The setting of the threshold level to avoid hallucinations is then a function of the ratio of the signal mean noise level to noise distribution width. This threshold is also dependent upon the number of messages *m* that are encoded (through equation 15)). Therefore a value of *m* corresponding to the worst case, determined from the lowest point on the curve in Figure 13 and denoted by *m'* can be used leading to:

$$\frac{E_t}{\mu} = \sqrt{2}\frac{\sigma}{\mu} erf^{-1}\left(1-2\left(\frac{1}{2^b - m'}\right)^{\frac{1}{N-a+1}}\right) + 1 \qquad 21)$$

With the current parameters the value *m'* =15, leading to a threshold value given by

$$E_t = 0.1\sigma + \mu \qquad 22)$$

## Extensions

Concurrent coding is efficient because messages with the same pre-fix will share marks in the codeword. This can be seen in the nonlinear relationship between the number of messages and the number of marks as shown in Figure 6. Whilst many messages can be overlaid into the codeword through a common hashing function, this does mean that any message can only appear once in any codeword, which could be inconvenient for practical applications. It is also not clear that any particular order for the messages can be maintained through the decoding process, thus the receiver is required to make sense of the decoded messages. What is needed is therefore some method for including within the codeword additional information about the contents. This could be done through selective use of control or handshaking messages. However another possibility is to encode several sets of messages using different hashing functions. This would allow a set of control messages to be overlaid upon the data messages. This would also allow the same messages to be included several times in the same codeword. This is a process similar to Interleaved division multiple access (IDMA)[19] where information from different users is randomly interleaved to prevent interference. This is of course a less efficient way of encoding the information as multiple use of marks is less likely and would increase the computational load in comparison to the same number of messages with one hash function. An initial trial of this approach was attempted using PRBS hash functions with different seeds. A set of messages was encoded several times using different hash states. It was found that with 2 hash states the message lists could be correctly decoded, each correctly decoding the same values. More hash states resulted in the generation of hallucinations which suggests there is a significant interaction between PRBS functions. Additionally the PRBS hash function was modified to provide several versions with different internal feedback connections representing different hash functions. This however generated hallucinations when more than one hash function was used. It seems clear that the principle is sound but requires the use of non-interacting hash functions to be effective. This should be the subject of future investigations.

## Conclusions

Concurrent codes offer a useful alternative to established encoding methods where robustness to noise and intermittency is required. This work has shown that a simple approach to keep down complexity and computational burden (by using small messages and a PRBS) can offer remarkable resilience to noise and intermittency. It has been shown that with contiguous chunks of missing data of a size up to 40% of the total transmitted data, the whole transmission can be recovered with perfect accuracy. This resilience to data intermittency can be achieved without the use of cyclic codes or interleaving. The use of indelible marks (an asymmetric binary code) means that the original data encoded into the transmission can always be recovered although sometimes – in cases of high noise or large sections of missing data - obscured by false decodings. Thus, even with in excess of 50% of the encoded transmission removed the original data was still received and decoded. This remarkable facility takes the use of concurrent codes well beyond the original remit of providing resistance to jamming without the use of a shared encryption key. This technique could be used in situations where it is vitally important that specific transmitted information is received, such as hostile military scenarios, or reliability in the conveyance of medical or security information. Information conveyance through harsh and noisy environments could be implemented. Concurrent coding has particular relevance in situations where transmissions are subject to random intermittency, such as free space optical communications where atmospheric scintillation can cause beam wander away from a receiver, or line of sight can be interrupted by moving vehicles. In such circumstances it is important to match the intermittency time to the codeword length to ensure correct reconstruction of the data. The same would also be true for rapid RF fading.

The requirement for the use of indelible marks aligns well with on-off keying modulation. Where power is only emitted for the 1's within the data, concurrent coding is significantly more efficient than other encoding schemes and would therefore be well suited to applications that require reduced power transmission, either to preserve stored power or to reduce the probability of signal interception.

Whilst it is certainly true that cyclic codes such as Reed Solomon encoding coupled with interleaving can offer a significant correction to burst errors and data corruption, there is value in an approach that is significantly simpler to understand and to implement. Concurrent codes are fundamentally different to cyclic codes and interleaving in the following ways: 1) They do not require additional data to be merged with the original through an encoding that records the parity. 2) Concurrent codes do not lose data when the corruption exceeds the capacity of the code to correct errors. 3) The decoding of concurrent codes is likely to be more computationally intensive than decoding of cyclic codes.

Clearly there is scope for more investigation into the design and application of concurrent codes with respect to code lengths, hashing functions, security concerns and interwoven data.

## References:

[1]. Martin Bossert.. *Channel Coding for Telecommunications* (1st ed.). John Wiley & Sons, Inc., New York, NY, USA(1999).

[2]. Glover, I., & Grant, P. M. Digital Communications, Pearson Education. (2010).

[3]. Berrou, C., Glavieux, A., & Thitimajshima, P. 'Near Shannon limit error-correcting coding and decoding: Turbo-codes.' 1. In Communications, 1993. ICC 93. Geneva. Technical Program, Conference Record, IEEE International Conference on , (1993). Vol. 2, pp. 1064-1070. IEEE.

[4]. Viterbi, A J "Error bounds for convolutional codes and an asymptotically optimum decoding algorithm". IEEE Transactions on Information Theory 13 (2): (1967). pp 260–269. doi:10.1109/TIT.1967.1054010

[5]. Lin, Shu, and Daniel Costello. "Error Control Coding: Fundamentals and Applications." ( Prentice Hall, 1983)

[6]. Clark Jr, George C., and J. Bibb Cain. Error-correction coding for digital communications. Springer Science & Business Media, 2013.

[7]. Fire, P. A class of multiple-error-correcting binary codes for non-independent errors, Sylvania Rept. RSL-E-2, Sylvania Reconnaissance Systems Laboratory, New York (1959)

[8]. http://www.usna.edu/Users/math/wdj/files/documents/reed-sol.htm, accessed January 2015.

[9]. Sklar B, 'Digital Communications: Fundamentals and Applications, Second Edition' (Prentice-Hall, 2001, ISBN 0-13-084788-7).

[10]. Wesolowski, Krzysztof. *Introduction to digital communication systems*. John Wiley & Sons, 2009.

[11]. Baird, Leemon C. III, Bahn, William L. & Collins, Michael D. 'Jam-Resistant Communication Without Shared Secrets Through the Use of Concurrent Codes', Technical Report, U. S. Air Force Academy, (2007).USAFA-TR-2007-01

[12]. Baird, Leemon C. III, Bahn, William L., Collins, Michael D., Carlisle, Martin C. & Butler, Sean 'Keyless jam resistance', Proceedings of the 8th Annual IEEE SMC Information Assurance Workshop (IAW), Orlando, Florida, June 20-22, (2007) pp 143-150. doi: 10.1109/IAW.2007.381926

[13]. Bahn, William L., Baird, Leemon C. III & Collins, Michael, D. 'Jam resistant communications without shared secrets', Proceedings of the 3rd International Conference on Information Warfare and Security (ICIW08), Omaha, Nebraska, (2008) April 24-25.

[14]. Baird, Leemon C. III, Carlisle, Martin & Bahn, William L 'Unkeyed Jam Resistance 300 Times Faster: The Inchworm Hash', MILCOM 2010 - Military Communications Conference, San Jose, CA, Oct(2010)

[15]. Baird, Leemon C III, Schweitzer, Dino, Bahn, William L & Sambasivam, Samuel. 'A Novel Visual Cryptography Coding System for Jam Resistant Communications', Journal of Issues in Informing Science and Information Technology, 7, (2010), pp 495-507,

[16]. Di Pietro, Roberto, and Gabriele Oligeri. "Jamming mitigation in cognitive radio networks." Network, IEEE 27.3 (2013).

[17]. http://www.dragonwins.com/domains/getteched/bbc/. Accessed 03/09/2015.

[18]. Sharma, P. S. (2012). Interleavers for IDMA Technology : A Comparison Survey, 1(2), 55–61.

[19]. Ping, L. P. L., Liu, L. L. L., Wu, K. W. K., & Leung, W. K. (2006). Interleave division multiple-access. IEEE Transactions on Wireless Communications, 5(4), 938–947. doi:10.1109/TWC.2006.1618943

[20]. Ping, L., Wang, P., & Wang, X. (2007). Recent progress in interleave-division multiple-access (IDMA). Proceedings - IEEE Military Communications Conference MILCOM, 1–7. doi:10.1109/MILCOM.2007.4455046

[21]. Baird, L. C., Carlisle, M. C., Bahn, W. L., & Smith, E. 'The Glowworm hash: Increased speed and security for BBC unkeyed jam resistance'. In military communications conference, 2012-milcom (2012, October). (pp. 1-6). IEEE.

**Appendix**

**Interleaving details for the Hamming Interleaved code.**

For concurrent coding the number of included messages can be variable and does not need to be known by the receiver to allow successful decoding. The only appropriate comparison for the HIC is to assume that the transmitter will interleave with a spacing given by the maximum available number of messages, and unused messages will just be zero padded. Hamming encoding of an 8 bit message gives a 16 bit result, thus there are 128 available messages and the codeword is interleaved into 16 sections. The Hamming encoding is capable of correcting 1 corrupted bit in each 4 bits of the original message. The original 8 bit message was broken into 2 chunks and the 16 bit Hamming encoded word created by appending the encoded bits. As interleaving spaces adjacent bits into neighbouring sections it would be possible for a gap of missing data greater than an interleaving spacing to corrupt 2 bits within a Hamming(8,4) encoding. Therefore the interleaving must be crossed between encodings such that an arrangement of word bits $a1a2...a16,b1b2...b16...$ is encoded as $a1b1c1...a8b8c8...a2b2c2..$. This ensures the maximum tolerance to data gaps is enacted. It follows that a missing section whose extent is larger than 2 interleaving distances will begin to corrupt the decoding, and an extent of 3 interleaving distances would usually corrupt all the messages. However in this implementation to ensure like-for-like comparison missing bits are replaced by zeros (as opposed to data erasure in some modulation schemes) which means that the scheme fails gracefully.